\documentclass[aps,preprint]{revtex4-1}
\usepackage{graphicx}
\usepackage{amsmath}
\usepackage{makeidx}
\usepackage{amsfonts}
\usepackage{amssymb}

\begin{document}

\title{Statistical physics approach to graphical games: local and global interactions}

\author{A. Ramezanpour}
\email{abolfazl.ramezanpour@polito.it}
\affiliation{Politecnico di Torino, C.so Duca degli Abruzzi 24, I-10129 Torino, Italy}

\author{J. Realpe-Gomez}
\email{john.realpe@polito.it}
\affiliation{Politecnico di Torino, C.so Duca degli Abruzzi 24, I-10129 Torino, Italy}

\author{R. Zecchina}
\email{riccardo.zecchina@polito.it}
\affiliation{Politecnico di Torino, C.so Duca degli Abruzzi 24, I-10129 Torino, Italy}
\affiliation{Collegio Carlo Alberto, Via Real Collegio 30, 10024 Moncalieri, Italy}

\date{\today}

\begin{abstract}
In a graphical game agents play with their neighbors on a graph to achieve an appropriate state of equilibrium. Here relevant problems are characterizing the equilibrium set and discovering efficient algorithms to find such an equilibrium (solution). We consider a representation of games that extends over graphical games to deal conveniently with both local a global interactions and use the cavity method of statistical physics to study the geometrical structure of the equilibria space. The method also provides a distributive and local algorithm to find an equilibrium. For simplicity we consider only pure Nash equilibria but the methods can as well be extended to deal with (approximated) mixed Nash equilirbia.
\end{abstract}

\pacs{} \maketitle

\section{Introduction}\label{S1}
In the last decade we observed a rapid merging of research interests in social sciences, economics and computer science driven in part by the common need of analyzing strategic interactions in multi-agent systems. There is a growing body of empirical work evidencing the influence of social interactions on economic outcomes which has encouraged economists to take explicit account of the {\it direct} (non-market) social influences on individual decision making \cite{BBDI-2010,DI-2010,O-2010}.

A general framework to study strategic interactions in multi-agent systems is game theory\cite{NM-1944}. It serves to analyze situations where self-interested agents, possibly with conflicting interests, struggle to get the best conditioned on (and conditioning) the behavior of others; the scope of the theory is to predict the strategic behavior that comes out in such situations, something that is usually done in the form of equilibrium concepts. Chief among them is the concept of Nash equilibrium (NE)  that is a strategy profile in which no agent has incentives to deviate {\it unilaterally }. Nash equilibria can be pure or mixed, depending on whether the agents are required to play deterministically or are allowed to randomize among their available strategies. While conceptually very compelling, the concept of pure NE has the drawback that it does not always exist, in contrast with the universality of the concept of mixed NE whose existence is guaranteed for any finite game \cite{N-1950,N-1951}.

An important concern for computer scientists was to understand if Nash equilibria are actually efficiently computable \cite{NRTV-2007}. Much work has been devoted at analyzing the computational complexity of finding a Nash equilibrium \cite{DP-2006,DP-2005,CD-2005,CD-2006}. This has been proved to be complete in a class called PPAD \cite{P-1994} which contains problems believed to be hard problems\cite{DGP-2009,D-2009}. Indeed finding a Nash equilibrium typically becomes NP-hard as soon as we require it to satisfy certain natural properties (e.g. optimizing social welfare) \cite{GZ-1989,CT-2008} or when we restrict it to pure strategies \cite{GGS-2003,JS-2010}.

Statistical physics has also contributed important insights and techniques to these fields \cite{MM-book-2009}. For instance, in computer science it has provided powerful heuristic algorithms and a better understanding of the onset of computational complexity. From a physical point of view, a game is regarded as a system of interacting agents where an appropriate energy function maps the Nash equilibria to the ground states of the system. Our understanding of these systems has considerably improved in recent years mainly due to the concepts and tools developed in the study of complex systems displaying glassy behaviors \cite{MP-2001,MPZ-2002,MP-2003,KMRSZ-2007}.

From an economics modeling perspective, an interesting aspect of graphical games is
to study the interplay between local and global interactions \cite{HS-2006,HS-appear,BHO-2006}. Motivated by this, we will study a global graphical game where agents, besides the local payoffs, receive some global payoffs depending on an aggregate quantity, here the average strategy of the game or global magnetization. We design a message passing algorithm with a polynomial time complexity which is exact when the graph of local interactions is a tree and show how to obtain an approximated algorithm to deal with the global interaction in a more efficient way. This algorithm resembles the two step strategy used by Horst and Scheinkman \cite{HS-2006} to prove the existence of equilibria in a class of multi-agent systems with local (pairwise) and global interactions. Namely, we consider the magnetization as a fixed parameter, rendering the game an effective local graphical game, and require consistency in that we force the average magnetization to be equal to the fixed parameter. We test all these algorithms in some ensembles of random graphical games with global interactions, focusing mainly on an extension of the best shot game or maximal independent set problem \cite{BK-2007,GGJVY-2008}. Here we present upper bounds for the entropy and for the probability of having solutions which along with the results obtained by the cavity method of statistical physics help us to characterize the solution space of the problem.

In this paper we take the constraint satisfaction approach and for the sake of simplicity we focus on pure Nash equilibria,  but the methods can properly be generalized to deal with mixed equilibria. We show numerically how the information contained in the Belief Propagation (BP) messages \cite{KFL-2001} can be exploited to turn the algorithm into a one phase, fully distributed (and typically efficient) solver which directly converges to a {\it single} Nash equilibrium. This class of message passing solvers are called reinforced belief propagation (rBP)\cite{CFMZ-2005,BZ-2006}. Here we will consider the ensemble of independent payoffs that could be fully random or with some hidden solutions; in both cases we find typical Nash equilibria, if they exist, in a replica symmetric phase and easy to discover with the rBP algorithm. However,  finding an optimal Nash equilibrium maximizing the total payoff would be still difficult as we enter into a replica symmetry broken phase with a more complex organization of the optimal solutions.

The paper is organized as follows. In section \ref{S2} we give some definitions that will be used in the paper. Section \ref{S3} includes our results on graphical games with local random payoffs which could have hidden solutions. In section \ref{S4} we study
graphical games with a global interaction that depends on the average strategy of the agents. The conclusion is given in section \ref{S5}. In appendix \ref{remarks} we present more details on some rigorous statements mentioned in the text.

\section{Related works and definitions}\label{S2}
Traditionally, a game is defined by assigning a number (i.e. a payoff) to each player for each possible configuration of the other players. The number of parameters involved in this representation grows exponentially with the number of agents in the game. Therefore, it is crucial to exploit additional structures that may be present in certain situations to find an efficient representation of the problem.
Graphical games model the very common situation where each agent interacts only with a small subset of the whole population (see \cite{NM-1992,M-2000,KLS-2001,KM-2001} for related works). The interaction structure is encoded in a graph where each vertex corresponds to an agent and a link between two agents indicates that the two player's payoffs depend on each other's choice of strategy.

In Ref. \cite{KLS-2001} the authors provide a dynamic algorithm to find Nash equilibria on trees, which was later extended to deal with loopy graphs \cite{OK-2002}. They also drew analogies with the belief propagation algorithm \cite{KFL-2001} which is equivalent to the replica symmetric approximation in the cavity method. In fact the first phase of their algorithm can be understood as an instance of Warning Propagation (WP) where the probabilistic information contained in a belief is projected onto a boolean variable. Such a simplification allows to keep things in the realm of integer arithmetic and, in the case of graphical games, to prove the algorithm convergence. Connections to constraint satisfaction problems and Markov random fields can be found in \cite{VK-2002,DP-Markov-2006}.

Consider $N$ players indexed by $i=1,\dots,N$ playing a game with given payoffs  $M_i(\sigma_i|\sigma_{\partial i})$. A player payoff or utility depends on her strategy $\sigma_i \in \Lambda_i$ and the set of strategies played by her neighbors in a dependency graph $\mathcal{G}$, i.e. $\sigma_{\partial i}$. We use $\partial i$ to denote the neighbor set of player $i$.  In the following we
consider binary strategies, that is $\Lambda_i=\{-1,+1\}$, and positive payoffs $M_i(\sigma_i|\sigma_{\partial i})\in [0,1]$.

A strategy profile or configuration $\underline{\sigma}^*$ is a pure Nash equilibrium if for each player $\sigma_i^*$ is the best response to the neighbors actions, i.e.

\begin{equation}
M_i(\sigma_i^*|\sigma_{\partial i}^*) = \max_{\sigma_i} M_i(\sigma_i|\sigma_{\partial i}^*).
\end{equation}

Not any game possesses a pure equilibrium, but Nash theorem ensures that any finite game admits a mixed equilibrium \cite{N-1951}. Consider stochastic players where player $i$ chooses strategy $\sigma_i=+1$ with probability $x_i\in [0,1]$ and the other strategy with the complement probability. A mixed profile $\underline{x}^*$ is called a mixed Nash equilibrium if

\begin{equation}
\langle  M_i(\sigma_i|\sigma_{\partial i}) \rangle_{\sigma_i,\sigma_{\partial i}} =  \max_{\sigma_i} \langle  M_i(\sigma_i|\sigma_{\partial i}) \rangle_{\sigma_{\partial i}} ,
\end{equation}

where the averages are taken over strategies with respect to $\underline{x}^*$.

In this paper we shall mainly consider pure Nash equilibria. In cases where such an equilibrium does not exist,
we ask for approximate solutions that satisfy Nash conditions within some tolerated errors \cite{NRTV-2007}.
A configuration $\underline{\sigma}^*$ is called an $\epsilon$-Nash equilibrium if

\begin{equation}
M_i(\sigma_i^*|\sigma_{\partial i}^*) \ge M_i(-\sigma_i^*|\sigma_{\partial i}^*)-\epsilon.
\end{equation}

where $\epsilon \ge 0$.

We shall represent the above problem as a constraint satisfaction problem where a strategy profile is a Nash equilibrium if and only if it satisfies all the constraints.
To this end, we define a constraint $I_i(\sigma_i|\sigma_{\partial i})$ for each player $i$ to check if strategy $\sigma_i$ is a best response or not. That is $I_i(\sigma_i|\sigma_{\partial i})=1$ if $\sigma_i$ is the best response to the neighborhood configuration $\sigma_{\partial i}$, otherwise it is zero.
Statistical properties of the solution space can be obtained by studying
the following partition function

\begin{equation}
Z=\sum_{\underline{\sigma}}  e^{\beta \sum_i M_i(\sigma_i|\sigma_{\partial i})} \prod_i  I_i(\sigma_i|\sigma_{\partial i}),
\end{equation}

where $\beta$ is a parameter to optimize over the solution space. Two special limits $\beta=0$ and $\beta \to \infty$ give the typical and optimal welfare solutions, respectively. 

The reader can find more about physical approaches to games in \cite{BE-1998,SF-2007} and references therein.

\section{Local graphical games}\label{S3}

We first study local graphical games where each player payoff depends only
on its local neighborhood on the graph.
In particular we consider random regular graphs ($|\partial i|=K$ for all players).
The aim is to characterize the equilibria space, for example the number of equilibria and their geometrical organization.
To do this, we resort to the cavity method of statistical physics in the replica symmetric (RS) and 1-step replica symmetry breaking (1RSB)
approximations \cite{MM-book-2009}.

\subsection{RS equations}\label{S31}
In the replica symmetric approximation we assume that all solutions belong to a single cluster in the configuration space; there is a path connecting any two solutions such that neighboring solutions along the path are different only on a sub-linear number of players. As a result, correlations are short ranged and it is usually computationally easy  to find a solution (a Nash equilibrium) to the problem.    

Assume that our dependency graph $\mathcal{G}$ is a tree and consider cavity graph
$\mathcal{G}_{i\to j}$ including all the nodes and edges connected to $j$ through $i$.
Define $\pi_{i \rightarrow j}(\sigma_i;\sigma_j)$ as the probability of having strategies
$\sigma_i$ and $\sigma_j$ in a solution of $\mathcal{G}_{i\to j}$ when constraint $I_j$ is ignored.
Then one can easily write an equation for $\pi_{i \rightarrow j}(\sigma_i;\sigma_j)$ relating it to other cavity probabilities \cite{MM-book-2009}:

\begin{equation}
\pi_{i \rightarrow j}(\sigma_i;\sigma_j) \propto  \sum_{\{\sigma_k|k \in \partial i\setminus j\}} e^{\beta M_i} I_i \prod_{k \in \partial i \setminus j}\pi_{k \rightarrow i}(\sigma_k;\sigma_i).
\end{equation}

These equations are called belief propagation equations and can be solved iteratively starting from random initial cavity probabilities or messages.
Having the cavity probabilities we can obtain the free energy by the Bethe expression:

\begin{equation}
F=\sum_i \Delta F_i- \sum_{(ij) \in \mathcal{G}} \Delta F_{ij},
\end{equation}

where $\Delta F_i$ and $\Delta F_{ij}$ are the free energy shifts by adding node $i$ and link $(ij)$, respectively.
In a tree graph one obtains:

\begin{eqnarray}
e^{-\beta \Delta F_i}=\sum_{\sigma_i}\sum_{\sigma_{\partial i}} e^{\beta M_i} I_i \prod_{j\in \partial i} \pi_{j \rightarrow i}(\sigma_j;\sigma_i),\\ \nonumber
e^{-\beta \Delta F_{ij}}=\sum_{\sigma_i,\sigma_j} \pi_{i \rightarrow j}(\sigma_i;\sigma_j)\pi_{j \rightarrow i}(\sigma_j;\sigma_i).
\end{eqnarray}

We expect the free energy to be asymptotically correct in locally tree graphs as long as correlations are short range, i.e. in a replica symmetric phase.

The BP equations can also be used as an algorithm to find a Nash equilibrium.
Reinforcement is a way of doing this by progressively biasing the players to take the strategy that is suggested by the BP marginals \cite{BZ-2006}. The reinforced BP equations are

\begin{eqnarray}
\pi_{i \rightarrow j}(\sigma_i;\sigma_j)\propto [\pi_{i}(\sigma_i)]^r
 \sum_{\{\sigma_k|k\in \partial i \setminus j\}}e^{\beta M_i} I_i \prod_{k \in \partial i \setminus j}\pi_{k \rightarrow i}(\sigma_k;\sigma_i),\\ \nonumber
\pi_{i}(\sigma_i)\propto [\pi_{i}(\sigma_i)]^r
\sum_{\{\sigma_{k}|k \in \partial i\}}e^{\beta M_i} I_i \prod_{k \in \partial i}\pi_{k \rightarrow i}(\sigma_k;\sigma_i),
\end{eqnarray}

where $r\ge 0$ is the reinforcement parameter. One can start with random initial values for the messages and update them according to the rBP equations in a random sequential way. At the beginning we set the reinforcement parameter to zero and increase its value slowly while the system converges to a solution. Notice that for $r\to \infty$ any solution of the problem is a fixed point of the above equations. The rBP equations thus suggest a local and distributive message passing algorithm to approach a Nash
equilibrium.

\subsection{1RSB equations}\label{S32}
For simplicity let us take the limit $\beta\to 0$ where $-\beta F$ is equivalent to entropy $S$.
In the one-step replica symmetry breaking framework we assume there exist an exponentially large number
of clusters of solutions  \cite{MM-book-2009}. This number is given by the so called complexity or configurational entropy by $e^{N\Sigma}$.
Clusters can have different  internal entropies (or sizes)  $s\equiv S/N$ and  $\Sigma(s)$ is used to indicate  the complexity of a cluster of size $s$. The so called dominant  clusters are those that maximize $\Sigma(s)+s$; with high probability a randomly selected solution belongs to these kind of clusters (though any specific algorithmic strategy would end up in clusters which are not necessarily the dominant ones).
It is useful to introduce Lagrange multiplier $m$ and work with generalized free energy $m\Phi=\Sigma(s)+ms$. This allows to obtain the complexity by an inverse Legendre transform after computing 
the generalized free energy. As long as we are in the RS phase (even if the solution space is clustered) the relevant clusters are those corresponding to $m=1$. This complexity goes continuously to zero at the thermodynamic RSB phase transition, after that the physical $m$ would be less than $1$, its value determined by the point of zero complexity.     

We assume that each BP fixed point corresponds to a cluster or state of the system. 
The 1RSB equations give the statistics of BP messages among different clusters \cite{MM-book-2009}

\begin{equation}
P_{i \rightarrow j}(\pi_{i\to j}) \propto \int \prod_{k\in \partial i\setminus j} dP_{k \rightarrow i}(\pi_{k \rightarrow i}) e^{m\Delta S_{i \rightarrow j}} \delta(\pi_{i\to j}-BP_{i \rightarrow j}),
\end{equation}

where

\begin{equation}
e^{\Delta S_{i \rightarrow j}}= \sum_{\sigma_i, \sigma_{\partial i}}  I_i \prod_{k\in \partial i \setminus j} \pi_{k \rightarrow i}(\sigma_k;\sigma_i).
\end{equation}

Here $m$ is called Parisi parameter and is to control the entropy density.
Given the cavity distributions $P_{i \rightarrow j}(\pi_{i\to j})$ we obtain the 1RSB free energy in the Bethe approximation

\begin{equation}
\Phi=\sum_i \Delta \Phi_i- \sum_{(ij) \in \mathcal{G}} \Delta \Phi_{ij},
\end{equation}

where

\begin{eqnarray}
e^{m\Delta \Phi_i}=\int \prod_{j\in \partial i} dP_{j \rightarrow i}(\pi_{j \rightarrow i})  e^{m\Delta S_{i}},\\ \nonumber
e^{m\Delta \Phi_{ij}}=\int dP_{i \rightarrow j}(\pi_{i \rightarrow j})dP_{j \rightarrow i}(\pi_{j \rightarrow i}) e^{m \Delta S_{ij}}.
\end{eqnarray}

The 1RSB free energy is related to the entropy and complexity by $m\Phi(m)=\Sigma(s)+m s$ where

\begin{eqnarray}
m=-\frac{\partial \Sigma(s)}{\partial s},\hskip1cm 
\Sigma(s)=-m^2 \frac{\partial \Phi(m)}{\partial m}.
\end{eqnarray}

The total 1RSB  entropy is given by $\Sigma(s)+s$ computed at the physical $m$. As long as we are in the RS
phase the 1RSB entropy computed at $m=1$ is equal to the BP entropy.

For a given graphical game we can solve the 1RSB equations with the population dynamics technique \cite{MP-2001,MP-2003}. We represent the probabilities $P_{i \rightarrow j}(\pi_{i\to j})$
on each directed edge of the graph with a population $Pop_{i \to j}$ of BP messages $\{\pi_a|a=1,\dots,N_p\}$. To update $Pop_{i\to j}$, we first
select randomly messages $\pi_{a_k}$ from $Pop_{k\to i}$ for $k\in \partial i\setminus j$.  Then the new message $\pi_{i \to j}$ and the entropy shift $\Delta S_{i \to j}$
are computed according to the BP equations, and with probability $\propto e^{m\Delta S_{i \to j}}$ a randomly selected message form $Pop_{i \to j}$ is replaced with
the new one. After converging to a stationary state we compute the average quantities by taking samples from the populations.

With the same scheme, one can write the 1RSB equations for other values of $\beta$, replacing the entropy with free energy.

\subsection{Random payoffs}\label{S33}

\begin{figure}
\includegraphics[width=10cm]{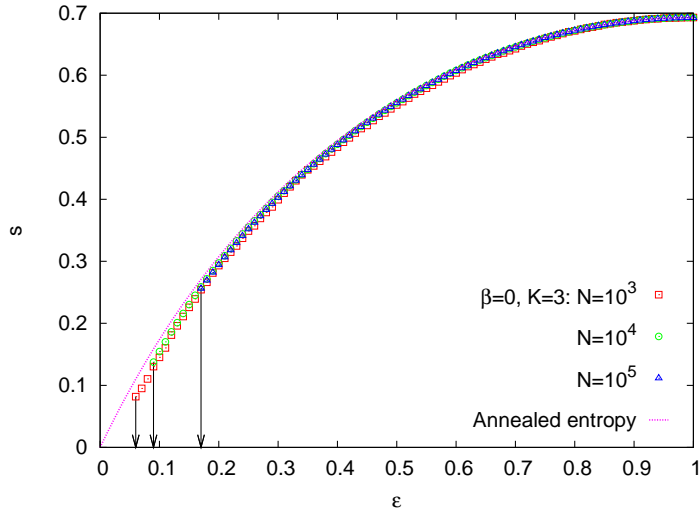}
\caption{BP entropy in the local graphical game with random payoffs.}\label{f1}
\end{figure}

Let us start with fully random payoffs where each element $M_i(\sigma_i|\sigma_{\partial i}) \in [0,1]$ is
a uniform random number independent of the other payoffs.
As figure \ref{f1} shows, here the BP entropy is positive only for approximate solutions with $\epsilon >\epsilon_c(N)$.
For smaller $\epsilon$ the BP algorithm finds contradictory messages, something that usually happens when
the solution set is empty. Moreover, the critical $\epsilon$ increases with $N$ and finally for $N\to \infty$ we would have $\epsilon_c=1$. 
Notice that for $\epsilon \ge 1$ any configuration is a solution to the problem.

It is easy to compute the average number of solutions or annealed entropy by averaging over the randomness in the payoffs

\begin{eqnarray}
s^{annealed}= \frac{1}{N} \ln \langle Z \rangle= \ln(2)+\ln \left(\frac{1+2 \epsilon -\epsilon^2}{2} \right).
\end{eqnarray}

The above annealed entropy,  displayed in figure \ref{f1}, provides an upper bound for the correct number of solutions. However, this entropy is
always greater than zero, except for exact solutions at $\epsilon=0$.
To understand why the approximate solutions do not survive in the thermodynamic limit we need to resort to another argument \cite{DDM-2008}:
Consider an arbitrary region $\Omega$ of the graph $\mathcal{G}$. Suppose that we fix the strategies of boundary players $\partial \Omega$ to $\sigma_{\partial \Omega}$.
Depending on the boundary state and payoffs, we may have no best response solution for the players in $\Omega$. And this could happen for all boundary configurations $\sigma_{\partial \Omega}$. We denote the probability of this event by $P_{no solution}(\Omega)$. 
Consider a collection  $\mathcal{C}$ of $N_{\Omega}$ disjoint regions in $\mathcal{G}$. Then, the probability of having a solution
is overestimated by

\begin{eqnarray}
P_{solution}\le \prod_l (1-P_{no solution}(\Omega_l)).
\end{eqnarray}

The simplest choice is when each region consists of a single player. Obviously in this case $P_{no solution}=0$, which leads to a trivial inequality for $P_{solution}$.
We can choose a larger region consisting of two neighboring players in the graph.  For uniform and independent random payoffs one finds $P_{no solution}(\Omega)\le (1/8)^{2^{|\partial \Omega|}}$.
As long as $|\partial \Omega|$ is finite and $N_{\Omega}=O(N)$ this results to an exponentially small probability of having a solution.

\begin{figure}
\includegraphics[width=7cm,angle=270]{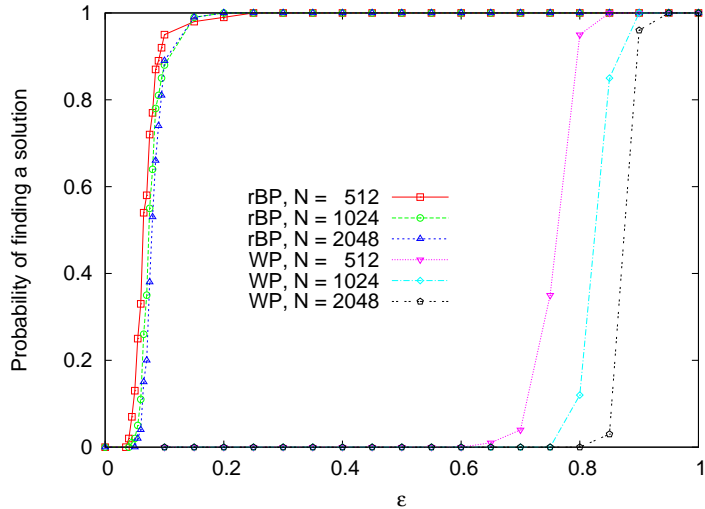}
\caption{Success probability of the rBP algorithm compared with the WP algorithm of Ref.\cite{KLS-2001} in the local graphical game with random payoffs.
The data have been obtained by running the algorithms on $100$ problem instances defined on random regular graphs of degree $K=3$.}\label{f2}
\end{figure}

Any way, given a finite game, we find that BP always converges as long as $\epsilon>\epsilon_c(N)$;
the system is in the RS phase with zero complexity and we can easily find an $\epsilon$-Nash equilibrium using our reinforced BP equations.
In figure \ref{f2} we compare the success probability of this algorithm with another message passing algorithm, similar to warning propagation,
introduced in Ref. \cite{KLS-2001}.  Here, a simple heuristic algorithm like Best Response (BR) does not converge to a solution.
In the BR algorithm, we start from an initial strategy profile and
as long as some players are not satisfied we randomly select one and update its strategy
to the best response.

As mentioned above the random payoff ensemble has a trivial thermodynamic limit. To get around this
problem we shall consider random games where nontrivial solutions exist.

\subsection{Random payoffs with hidden solutions}\label{S34}

We can always modify random payoffs to ensure that our game has at least some pure Nash equilibria.
Suppose that we want configuration $\underline{\sigma}^*$ to be a solution to the problem.
Then we modify the payoffs in the following way: for each player $i$, if necessary, we swap the two values $M_i(\sigma_i^*|\sigma_{\partial i}^*)$
and $M_i(-\sigma_i^*|\sigma_{\partial i}^*)$ to satisfy the Nash condition for the player.

\begin{figure}
\includegraphics[width=10cm]{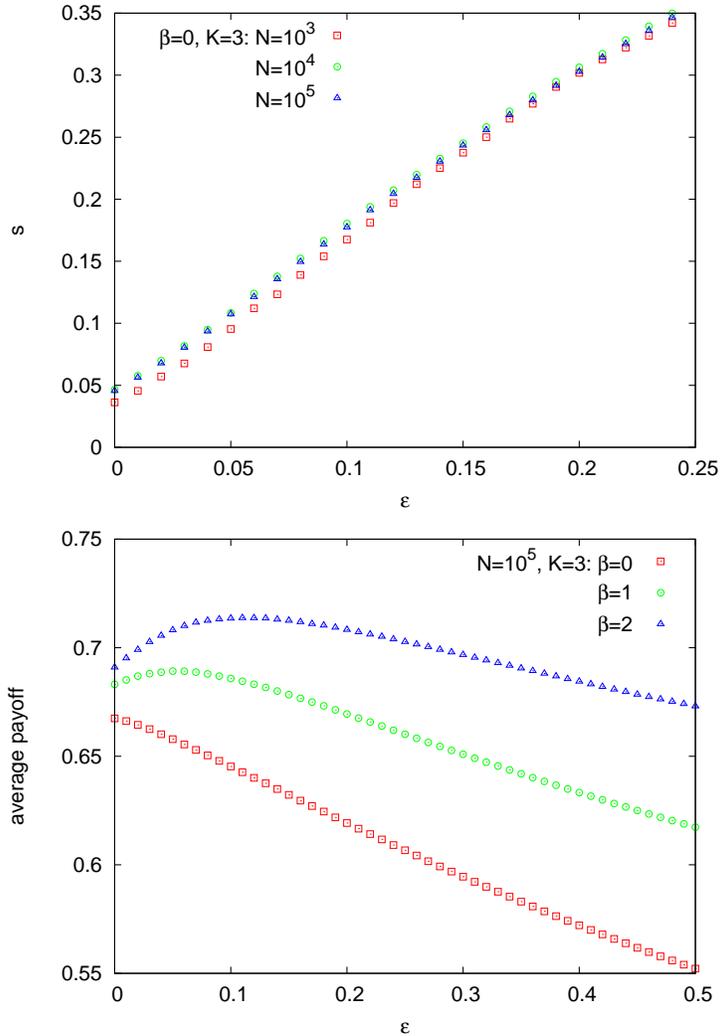}
\caption{BP entropy and average payoff in the local graphical game with a hidden solution.}\label{f3}
\end{figure}

Let us take the uniform and independent random payoffs to see how the picture changes when we plant random configuration $\underline{\sigma}^*$
into the solution space. For each player we choose $\sigma_i^*$ with equal probability from $\{-1,+1\}$.
Figure \ref{f3} displays the BP entropy computed on some large instances of the problem.
Interestingly, planting only one solution is enough to have an exponential number
of pure Nash equilibria at $\epsilon=0$. Moreover, these solutions do not disappear in the large $N$ limit,
as it happens for games with random payoff. To understand this consider the planted solution and a pair of neighboring players.
There is a finite probability that after flipping the two strategies we get another Nash equilibrium. Moreover, there is an extensive 
number of independent such players that could result to an exponential number of solutions close to the planted one.

\begin{figure}
\includegraphics[width=10cm]{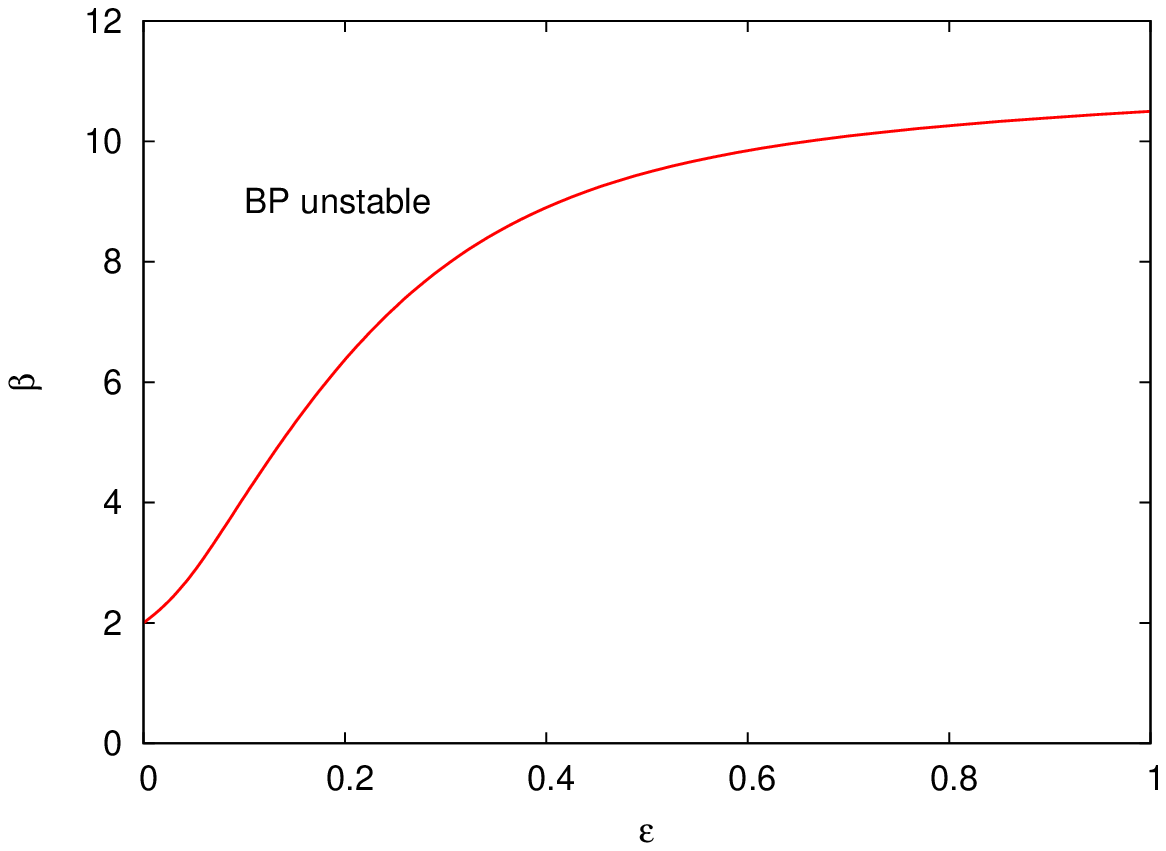}
\caption{The phase diagram in the local graphical game with a hidden solution.
The line has been obtained by checking the convergence of BP equations on a single instance of size $N=10^5$ and degree $K=3$. For $\beta>\beta_c$ the equations do not
converge in $T_{max}=1000$ iterations.}\label{f4}
\end{figure}

\begin{figure}
\includegraphics[width=10cm]{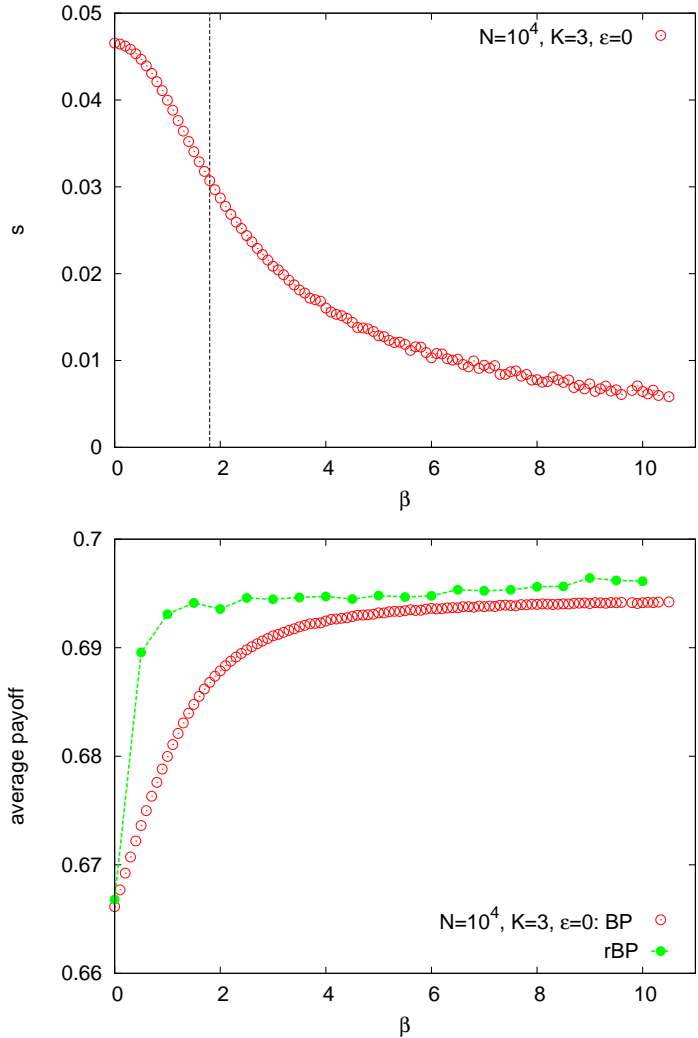}
\caption{BP entropy and average payoff in the local graphical game with a hidden solution.
The vertical line separates the RS (small $\beta$) and RSB (large $\beta$) regions. 
The rBP average payoffs are results of averaging over at least $50$ solutions obtained with the rBP algorithm.}\label{f5}
\end{figure}

Introducing temperature into the problem makes the phase diagram more interesting in that we observe
a critical line $\beta_c(\epsilon)$ separating the RS and RSB phases in the $\beta-\epsilon$ plane, see figure \ref{f4}; BP does not converge for $\beta>\beta_c$, signaling an RSB  phase transition. Moreover, there is a finite temperature-gap for any $\epsilon$, meaning that finding a typical $\epsilon$-Nash equilibrium is an easy task.
Figure \ref{f5} shows how the entropy and average payoff change with $\beta$ for exact Nash equilibria. For comparison we have also given the average payoff of
solutions obtained with the rBP algorithm.   
In figure \ref{f3} we also display the average total payoff as a function of $\epsilon$ for different values of $\beta$.
We observe a maximum appearing in the average payoff as we increase $\beta$;
obviously for small $\beta$, more accurate solutions have larger total payoff than solutions satisfying Nash condition at a larger $\epsilon$, since Nash condition is already maximizing the local payoffs. However, when $\beta$ is large enough we select those strategies maximizing the total payoff and with a larger $\epsilon$ we have more space to find a better global maximum.

\begin{figure}
\includegraphics[width=10cm]{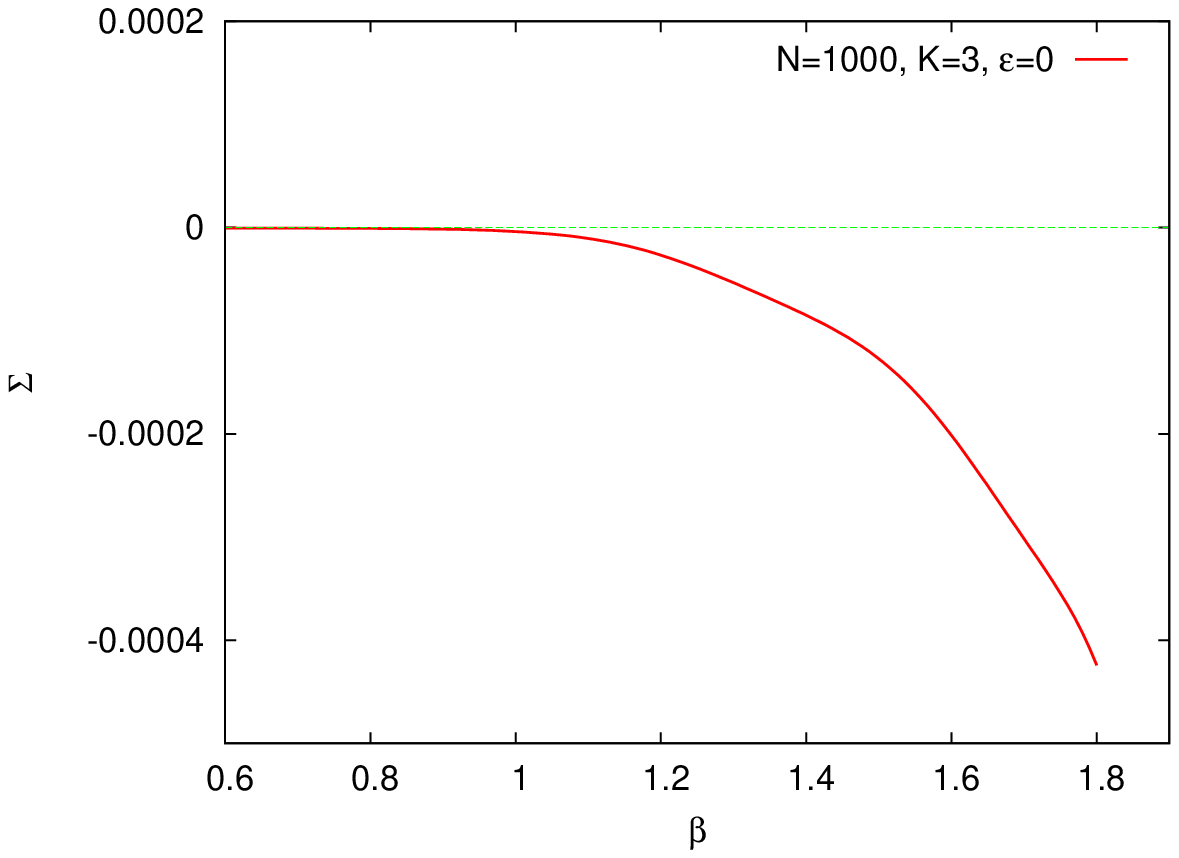}
\caption{$m=1$ complexity in the local graphical game with a hidden solution. The data are obtained by solving 1RSB equations with population dynamics
on a single instance of the problem. On each directed edge of the graph we have a population of size $1000$.}\label{f6}
\end{figure}

A more accurate estimate of the entropy and of the total payoff is obtained by considering replica symmetry breaking.
Figure \ref{f6} shows the $m=1$ complexity computed in the 1RSB
approximation at different temperatures.  The system is in the RS phase for small $\beta$;
BP converges, complexity is zero and the total 1RSB entropy is equal to the BP entropy. For larger $\beta$ we have
replica symmetry breaking, more precisely we enter into a condensed phase where only a finite number of solution clusters are relevant.
In this case, as we see in the figure, the $m=1$ complexity is negative but the relevant $m$ will be less than $1$ where complexity is zero.

\begin{figure}
\includegraphics[width=10cm]{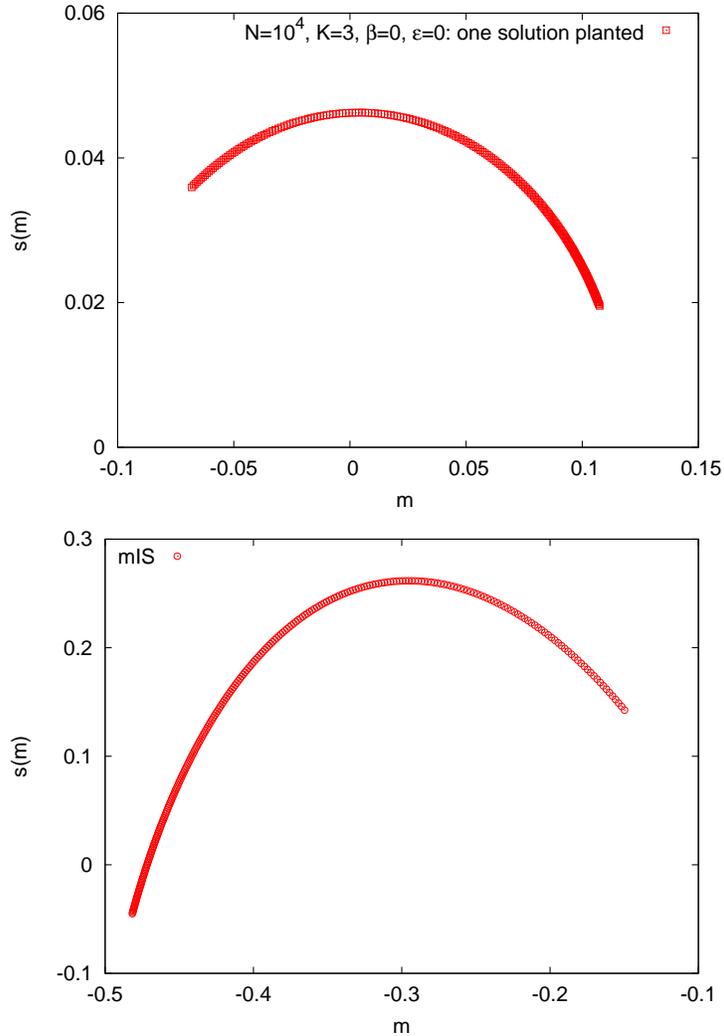}
\caption{Entropy as a function of magnetization in the local graphical game with hidden solutions. We plot the entropy only if the BP fixed point is stable.}\label{f7}
\end{figure}

Another way of planting solutions is to modify the random payoffs to mimic the constraints in an already known problem.
In this case, we go through all the neighboring configurations of a player and if necessary
swap the two values $M_i(\sigma_i|\sigma_{\partial i})$ and $M_i(-\sigma_i|\sigma_{\partial i})$ according to the constraints.
An example that we will later study in this paper is the maximal independent set (mIS) problem where a
player plays $+1$ only if all its neighbors play $-1$. Regarding the payoffs, it means that for each player
we need to have $M_i(+1| \forall j \in \partial i, \hskip2mm \sigma_j=-1) \ge M_i(-1| \forall j \in \partial i, \hskip2mm \sigma_j= -1)$ and
$M_i(-1| \exists j \in \partial i, \hskip2mm \sigma_j= +1) \ge M_i(+1| \exists j \in \partial i, \hskip2mm \sigma_j= +1)$.
It is easy to obtain a typical solution for this problem by running the Best Response algorithm.
What is difficult is to find an optimal solution, for example with a large number of active players.
Algorithms based on the BP equations help us to find good optimal solutions in large problem instances \cite{DPR-2009}.
In figure \ref{f7} we show the entropy of solutions with magnetization $m=\frac{1}{N} (\sum_i \sigma_i)$ in the region that BP equations converge.

\section{Global graphical games}\label{S4}

In a global graphical game the payoffs depend, besides the local neighborhood,
on the global state of the system, for instance as

\begin{equation}
M_i= M_i^{local}(\sigma_i|\sigma_{\partial i})+ M_i^{global}(\sigma_i|g),
\end{equation}

where $g(\underline{\sigma})$ is an aggregate quantity depending on the strategy profile.
As for local games the total number of solutions can be written as

\begin{equation}
Z=\sum_{\underline{\sigma}}  \prod_i I_i(\sigma_i|\sigma_{\partial i},g),
\end{equation}

where $I_i$ is an indicator function to check Nash condition for player $i$.
An interesting example is

\begin{equation}
M_i^{global}=h\sigma_i m, \hskip1cm m=\frac{1}{N}\sum_i \sigma_i,
\end{equation}

where players will receive more than their local payoffs if they are in majority or minority, depending on the sign of global field $h$.
These kind of interactions are similar to mean field models in statistical physics but in a different setting.
It also makes sense from a social point of view to have such global incentives \cite{OT-2009}.

The sole global problem $M_i=h\sigma_i m$ has only two solutions with magnetizations $m=\pm 1$ for $h>0$ and $C(N,N/2)$ solutions
with magnetization $m=0$ when $h<0$ and $N$ is even. Here $C(N,l)=\left(\begin{array}{c}
N\\
l
\end{array}\right)$ is the Binomial coefficient.

One could consider other global terms more suited to the local problem. For instance, we might be interested on the
total parity of solutions, where payoffs are given by

\begin{equation}
M_i= M_i^{local}(\sigma_i|\sigma_{\partial i})+ h \prod_i \sigma_i.
\end{equation}

In statistical physics these type of interactions are studied to model structural glasses.
In this case the global problem $M_i=h \prod_i \sigma_i$ partitions the configuration space to even and odd parity solutions for
$h>0$ and $h<0$, respectively.

In this paper we shall study the former problem where total activity or magnetization $m$
determines the global payoff. For the sake of simplicity, in the rest of the paper we shall work in the replica symmetric approximation and set $\beta=0$ and $\epsilon=0$.

\subsection{Rigorous results}\label{S41}

Suppose that local payoffs take integer values in $\{0,1\}$ and there is no degeneracy, that is
$M_i^{local}(\sigma_i|\sigma_{\partial i})\ne M_i^{local}(-\sigma_i|\sigma_{\partial i})$. This special case
helps us to understand the problem better when studying real and random payoffs.
Consider a strategy profile $\underline{\sigma}$ with magnetization $m$.
Flipping the strategy $\sigma_i$ of player $i$ results to the following change in her  payoff:

\begin{eqnarray}
\Delta M_i=\Delta M_i^{local}-2h(\sigma_i m-\frac{1}{N}),
\end{eqnarray}

where the $1/N$ term comes from the change in total magnetization. In order to simplify the arguments, we will consider only strict solutions, 
i.e. those with $\Delta M_i < 0$.
We say a player is locally happy if $\Delta M_i^{local}<0$ and group the players in distinct sets $H_+, H_-, U_+, U_-$ for
locally happy ($H$) and unhappy ($U$) players with plus and minus strategies. It is easy to see that players in $H_+$ satisfy the Nash condition
only if:

\begin{equation}
\left\{ \begin{array}{ll}
    m-\frac{1}{N} < \frac{1}{2|h|}, & \hbox{$h<0$;} \\
    m-\frac{1}{N} > -\frac{1}{2|h|}, & \hbox{$h>0$;} \\
  \end{array}
\right.
\end{equation}

The above conditions define two lines in the $m-h$ plane.
Similarly we can write the conditions for the other sets. From these we conclude that:

\textbf{Remark (1)}:
For $-\frac{1}{2(1-1/N)}<h<\frac{1}{2(1+1/N)}$ any local solution is also a solution of the problem. A local solution is a solution of type $(H_+,H_-)$.
The entropy of local solutions survived at
global field $h$ is:
\begin{eqnarray}
s^{local}(h)=\max_{-m^*(h)<m<+m^*(h)} s^{local}(m),
\end{eqnarray}

where
\begin{eqnarray}
m^*(h)=\min \left(1,\frac{1}{2|h|}-\frac{\mathrm{sgn}(h)}{N}\right).
\end{eqnarray}

We defined $s^{local}(m)$ as the entropy of local solutions (i.e. at $h=0$) having magnetization $m$.

\textbf{Remark (2)}:
For $h<-N/2$ and $h>N/2$ only global solutions remain. For positive $h$ the two global solutions with $m=\pm 1$ appear at $h=\frac{1}{2(1-1/N)}$,
whereas for negative $h$ the global solutions with $m=0$ appear at $h=-N/2$.

\textbf{Remark (3)}:
For $-N/2<h<N/2$ one may have mixed solutions. These are solutions of type $(H_+, H_-, U_+)$ (for $hm>0$) or $(H_+, H_-, U_-)$ (for $hm<0$).
That is we can not have solutions with both sets $U_+$ and $U_-$ non-empty. There is no mixed solution for $h>0$, and if there is a mixed solution for $h<-1/2$ it must have magnetization $m=\pm \frac{1}{2|h|}$.

When the local payoffs are uniform random numbers in $[0,1]$ satisfying the maximal independent set constraints, we have:

\textbf{Remark (4)}:
There is no solution with magnetization $m\in ]2\rho_{max}-1,0[$ for $h>0$, where $\rho_{max}$ denotes the size of the maximum independent set in graph $\mathcal{G}$.
Moreover, the probability of having a solution is zero in the thermodynamic limit when $0<2h(m \pm 1/N)<1$.

The reader can find more about these remarks in appendix \ref{remarks}.

\subsection{Annealed approximation}\label{S42}

Given the ensemble of local payoffs we can compute the average number of solutions in
a graphical game as

\begin{eqnarray}
\langle Z \rangle=\langle e^{Ns} \rangle=\sum_{\underline{\sigma}}\prod_i \langle I_i \rangle.
\end{eqnarray}

The average is taken over the randomness in
the independent local payoffs. The convexity of exponential function ensures that $s^{annealed}\equiv \ln \langle Z \rangle$
is an overestimate of the average entropy $\langle \ln Z \rangle$. In a global graphical game the constraint $I_i$ depends also on total
magnetization density $m$, therefore, we will restrict the above equations to the subspace of fixed $M=\sum_i \sigma_i$. Then the average number of solutions reads

\begin{eqnarray}
\langle Z \rangle= \sum_M \sum_{N(H_+),N(H_-)} e^{Ns[m,n(H_+),n(H_-)]} \prod_{\sigma=+1,-1} p(H_{\sigma})^{N(H_{\sigma})}p(U_{\sigma})^{N(U_{\sigma})},
\end{eqnarray}

where $e^{Ns[m,n(H_+),n(H_-)]}$ is the number of configurations with specified densities, e.g. $n(H_+)=N(H_+)/N$. Notice that the other two densities are not independent but
given by

\begin{eqnarray}
N(U_+)=\frac{N+M}{2}-N(H_+), \\ \nonumber
N(U_-)=\frac{N-M}{2}-N(H_-).
\end{eqnarray}

Moreover, as described in the previous section we can not have all densities nonzero; depending on the sign of $hm$ one of the two quantities
$N(U_+), N(U_-)$ should be zero, such that at the end there remains only one independent parameter $N(H_+)$ or $N(H_-)$.
For given $h$ and $m$, the probability that a locally happy player is satisfied by the total payoff is

\begin{eqnarray}
p(H_-)=\Pr(\Delta M^{local} > 2h(m+1/N) ), \\ \nonumber
p(H_+)=\Pr(\Delta M^{local} > -2h(m-1/N) ).
\end{eqnarray}

And if the player is locally unhappy

\begin{eqnarray}
p(U_-)=\Pr(\Delta M^{local} < -2h(m+1/N) ), \\ \nonumber
p(U_+)=\Pr(\Delta M^{local} < 2h(m-1/N) ).
\end{eqnarray}

For uniform random payoffs in $[0,1]$ we have

\begin{eqnarray}
\Pr(|\Delta M^{local}|>\epsilon)=\left\{ \begin{array}{ll}
    1  & \hbox{$\epsilon<0$;} \\
    (1-\epsilon)^2  & \hbox{$0\le \epsilon \le 1$;} \\
    0  & \hbox{$\epsilon>1$.}
  \end{array}
\right.
\end{eqnarray}

The above quantities are enough to compute the average number of solutions for random payoffs.
Let us separate the two cases of positive and negative fields.
For $h>0:$

\begin{equation}
\langle Z \rangle=  \frac{1}{2^N}\left\{ \begin{array}{ll}
    C(N,(N-M)/2)  p(H_+)^{(N+M)/2}[1+p(U_-)]^{(N-M)/2} & \hbox{$M < -1$;} \\
    C(N,(N-M)/2)  p(H_+)^{(N+M)/2} & \hbox{$M =-1$;} \\
    C(N,N/2)      p(H_-)^{N/2}p(H_+)^{N/2} & \hbox{$M =0$;} \\
    C(N,(N+M)/2)  p(H_-)^{(N-M)/2} & \hbox{$M =+1$;} \\
    C(N,(N+M)/2)  p(H_-)^{(N-M)/2}[1+p(U_+)]^{(N+M)/2} & \hbox{$M >+1$.}
  \end{array}
\right.
\end{equation}

And for $h<0:$

\begin{equation}
\langle Z \rangle= \frac{1}{2^N} \left\{ \begin{array}{ll}
    C(N,(N+M)/2)  p(H_-)^{(N-M)/2}[1+p(U_+)]^{(N+M)/2} & \hbox{$M < 0$;} \\
    C(N,N/2)  [1+p(U_-)]^{N/2} [1+p(U_+)]^{N/2}    & \hbox{$M =0$;} \\
    C(N,(N-M)/2)  p(H_+)^{(N+M)/2}[1+p(U_-)]^{(N-M)/2} & \hbox{$M >0$;} \\
  \end{array}
\right.
\end{equation}

Notice that given $h$ the entropy is symmetric with respect to $m$.
It is easy to see that the only positive contribution to the entropy comes from $m=0$ configurations when $h<0$ and scales with $N$:

\begin{eqnarray}
s^{annealed}=\ln \left[2-\left(1-\frac{2|h|}{N}\right)^2 \right].
\end{eqnarray}

In the other cases, the annealed entropy as a function of magnetization is always less than or equal to zero. Consider for example the case $h>0$ such that $0<2hm<1$, we have

\begin{eqnarray}
s^{annealed}=-\ln(2)-\left(\frac{1+m}{2}\right) \ln \left(\frac{1+m}{2}\right)-\left(\frac{1-m}{2}\right) \ln \left (\frac{1-m}{2} \right) \\ \nonumber
+\left(\frac{1+m}{2}\right)\ln[1+4hm-4h^2m^2]+\left(\frac{1-m}{2}\right) \ln[1-4hm+4h^2m^2].
\end{eqnarray}

The above entropy is zero only at $m=0$, or at $m=m_0(h)$ if $h$ is greater than critical value $h_c \approx 0.25$ determined by the following equation

\begin{eqnarray}
m_0=\tanh(h_g-h_l), \\ \nonumber
h_l=2h(1-m_0)\frac{1-2hm_0}{1-4hm_0+4h^2m_0^2} +\frac{1}{2}\ln[1-4hm_0+4h^2m_0^2], \\ \nonumber
h_g=2h(1+m_0)\frac{1-2hm_0}{1+4hm_0-4h^2m_0^2} +\frac{1}{2}\ln[1+4hm_0-4h^2m_0^2].
\end{eqnarray}

The nontrivial magnetization $m_0$ approaches to $1$ as global field $h$ reaches the value $1/2$.

The situation is a bit more complex when the local payoffs have some structure.
The difficulty comes from computing $s[m,n(H_+),n(H_-)]$ which was trivial for random payoffs.
When dependency graph $\mathcal{G}$ is a chain we can compute this entropy exactly. However, for arbitrary
graphs we will estimate it in the Bethe approximation by solving the following problem:

\begin{eqnarray}
Z(x,y)=\sum_{\underline{\sigma}} e^{x N(H_+)+y N(H_-)} I(U_{\tau} =\emptyset).
\end{eqnarray}

\begin{figure}
\includegraphics[width=10cm]{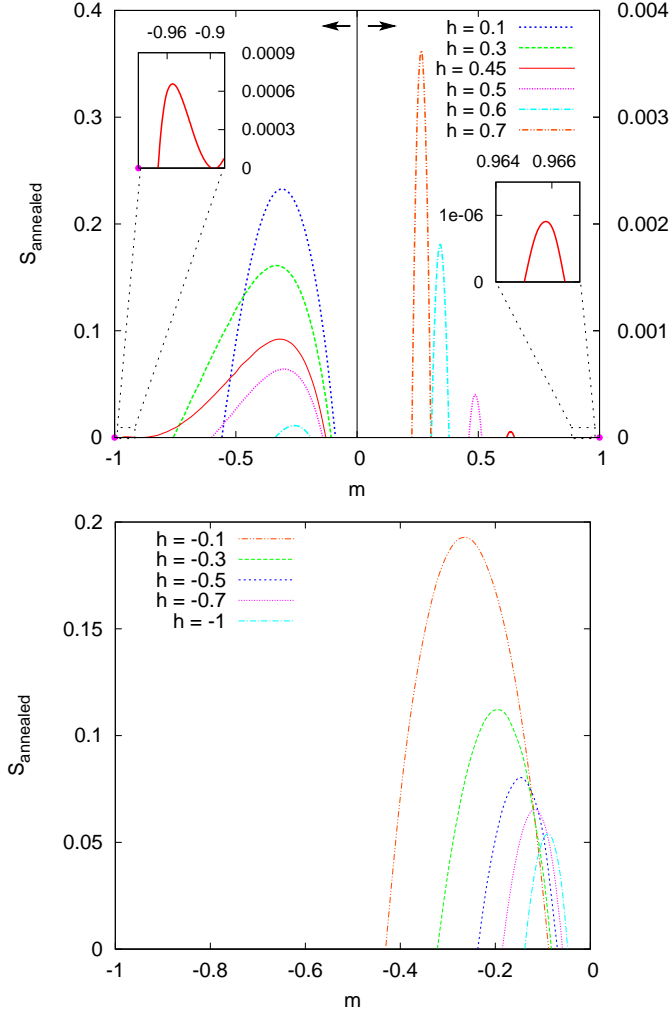}
\caption{Annealed entropy as a function of magnetization in the global problem with local mIS constraints. The entropy has been computed for random regular graphs of degree $K=3$.}\label{f8}
\end{figure}

Here $\tau=-\mathrm{sgn}(hm)$ and $I(U_{\tau} =\emptyset)$ is an indicator function to have $U_{+}$ or $U_{-}$ empty, depending on the sign of $hm$.
The entropy is obtained by a Legendre transformation after computing the free energy of this problem for appropriate values of fields $x$ and $y$.
Figure \ref{f8} shows the annealed entropy obtained in this way for random regular graphs of degree $K=3$ with local mIS constraints.
Notice the small entropy maximums appearing close to the global polarized solutions when $h$ is approaching $h_c\approx 0.43$.
We remind that according to remark (4) the entropy is zero in thermodynamic limit for $0<2h(m\pm 1/N)<1$. As for random local games in section \ref{S33},
here the annealed entropy does not give the correct behavior. However, it is still useful in that we obtain a qualitative picture of the entropy in finite size systems.

\subsection{Global algorithms}\label{S43}
The global graphical game can be treated like the local one by introducing a global constraint fixing the global quantity. Suppose that the global payoff depends on a global variable $g \in \Lambda_g$. We write the following partition function to count the
Nash equilibria

\begin{eqnarray}
Z=\sum_g \sum_{\underline{\sigma}} I_g(\underline{\sigma}) \prod_i I_i (\sigma_i|\sigma_{\partial i},g),
\end{eqnarray}

where as before $I_i$ checks for Nash condition and $I_g$ is an indicator function to fix quantity $g=g(\underline{\sigma})$.
We can then write the standard BP equations regarding $I_g$ as another constraint in the problem. However, in this way we
introduce a large number of loops into the problem, which destroys the BP exactness even when the original graph $\mathcal{G}$ is a tree.
To preserve this property here we follow another strategy in which the global constraint is broken to local ones in the expense of
introducing new variables. Consider global quantities that their computation can be partitioned into smaller local computations. This is the case for example when $g$ is total magnetization or parity. Then we introduce cavity variables $g_{i \to j}$ which are passed
along a spanning tree $\mathsf{T}$ of graph $\mathcal{G}$. These variables are determined by other cavity variables as $g_{i \to j}=g(\sigma_i,\{g_{k\to i}|k \in \partial i \setminus j, \mathsf{T} \})$. For instance, when $g$ is total magnetization, the cavity magnetizations are given by $g_{i\to j}=\sigma_i+\sum_{k\in \partial i \setminus j,\mathsf{T}} g_{k\to i}$. Now the global partition function can be rewritten as

\begin{eqnarray}
Z=\sum_{\{g_{i\to j}| (ij) \in \mathsf{T} \}} \sum_{\underline{\sigma}} \prod_i I_{ig} (\sigma_i|\sigma_{\partial i},g_i),
\end{eqnarray}

where $I_{ig}$ is to check the Nash condition and outgoing cavity variables $\{g_{i\to j}|j \in \partial i, \mathsf{T}\}$. Notice that each player computes its estimate of the global quantity 
$g_i=g(\sigma_i,\{g_{k\to i}|k \in \partial i,\mathsf{T}\})$ locally after receiving the incoming cavity variables. The BP equations with new variables read

\begin{equation}
\pi_{i \rightarrow j}(\sigma_i,g_{i \rightarrow j};\sigma_j,g_{j \rightarrow i})\propto  \sum_{\{\sigma_{k},g_{k\rightarrow i}| k \in \partial i \setminus j\}}
I_{ig} (\sigma_i|\sigma_{\partial i},g_i) \prod_{k \in \partial i\setminus j}\pi_{k \rightarrow i}(\sigma_k,g_{k \rightarrow i};\sigma_i,g_{i \rightarrow k}).
\end{equation}

The time complexity of this algorithm is $N K|\Lambda_g|^{k_{max}}$  where $k_{max}$ is the maximum degree in spanning tree $\mathsf{T}$.
Actually, the complexity can be reduced to $N K|\Lambda_g|^2$ if we pass the messages $g_{i\to j}$ along a spanning chain which is used just to compute the global
quantity.  Nevertheless, the algorithm is still computationally expensive specially when $|\Lambda_g|$ is large.

\begin{figure}
\includegraphics[width=10cm]{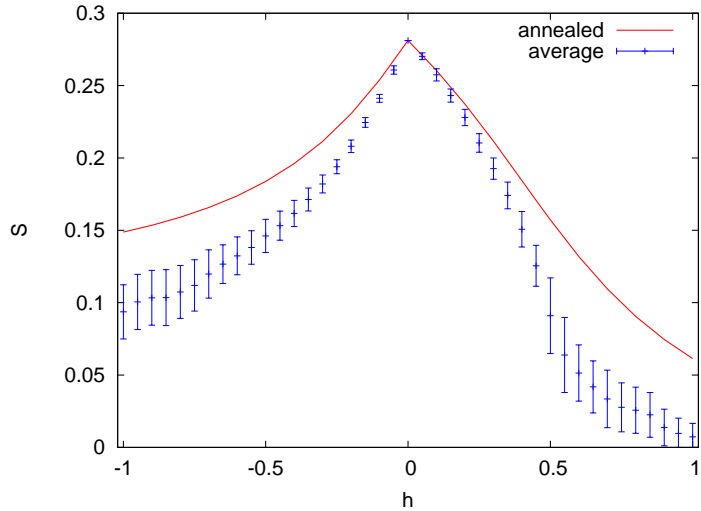}
\caption{Comparing the exact entropy with the annealed one for a chain of players in the global problem with local mIS constraints. The data are results of averaging over $10$ instances of random payoffs. }\label{f9}
\end{figure}

In figure \ref{f9} we compare the exact BP entropy computed in this way with the annealed entropy in a small chain of players, when $g=m$ and  $|\Lambda_g|=N$.

\subsection{Local algorithms for global games}\label{S44}
In this section we present another way of dealing with the global payoff when it depends on the   
total magnetization. Considering $h$ and $m$ as fixed parameters in the payoffs, we reduce the problem to  a local one but with modified conditions for solutions. The partition function for this local problem reads

\begin{equation}
Z(m)=\sum_{\underline{\sigma}} \prod_i I_i(\sigma_i|\sigma_{\partial i},m).
\end{equation}

Clearly the entropy computed in this way is an upper bound for $s(m)$, the entropy of solutions with
magnetization $m$ in the global problem. The reason is that here $m$ is just a parameter which is not necessarily the total magnetization.  
Indeed, we can do better than this by introducing an external field to really fix the total magnetization to $m$

\begin{equation}
Z(m)=\sum_{\underline{\sigma}} e^{x \sum_i \sigma_i} \prod_i  I_i(\sigma_i|\sigma_{\partial i},m),
\end{equation}

where $x$ is chosen such that

\begin{equation}
m=\frac{1}{N} \frac{\partial \ln Z(m)}{\partial x}.
\end{equation}

In the Bethe approximation the average magnetization is computed from the cavity messages satisfying BP equations

\begin{eqnarray}
\pi_{i \rightarrow j}(\sigma_i;\sigma_j)\propto
\sum_{\{\sigma_k|k\in \partial i \setminus j\}} e^{x \sigma_i} I_i(\sigma_i|\sigma_{\partial i},m) \prod_{k \in \partial i \setminus j}\pi_{k \rightarrow i}(\sigma_k;\sigma_i).
\end{eqnarray}

\begin{figure}
\includegraphics[width=10cm]{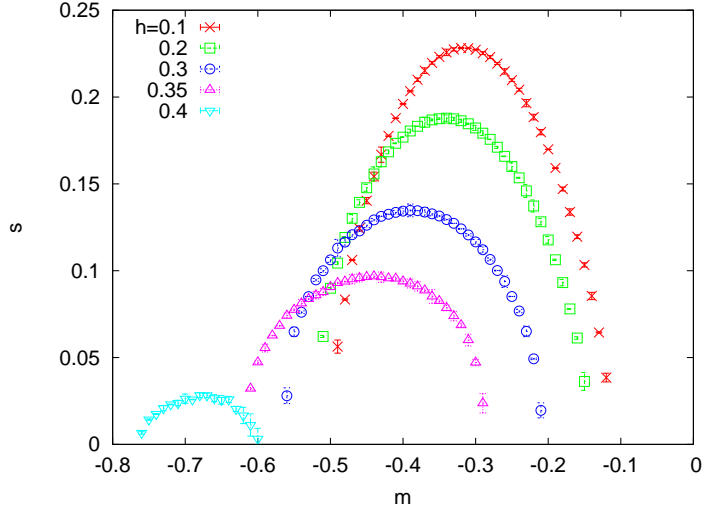}
\caption{BP entropy as a function of magnetization in the global problem with local mIS constraints. The data have been obtained with population dynamics with $N_p=2\times 10^5$. Here the errorbars show the difference between the fixed and converged magnetizations.} \label{f10}
\end{figure}

In this way we obtain a better estimation of $s(m)$ as displayed in figure \ref{f10}. 
This method of fixing magnetization has already been implemented in Refs.\cite{DMU-2004,MM-2006,SZ-2010}.

The above entropy can also be computed with the population dynamics technique. To do this we represent the set of BP messages in a graph by a large population of messages. The
population is updated by selecting randomly $K-1$ messages $\pi_{a_l}$ and computing a new BP messages according to the BP equations. Then a randomly selected element
of population is replaced with the new message. To fix the magnetization we also update field $x$ such that the expected magnetization in population is equal to $m$.

There are some subtle points here to mention  about the population dynamics.
First note that in the population dynamics we do not have the small payoff change $2h/N$ in the Nash conditions; we are at the thermodynamic limit $N\to \infty$, and $h$ is finite.
In all the numerical simulations we will work with a global field of order $1$, which is reasonable since the local payoffs are real numbers
in $[0,1]$. At the first sight the small term  $2h/N$ seems irrelevant for
large problem sizes but adding this correction could eliminate some solutions when $h$ is positive.
We recall that the Nash condition for player $i$ is $\Delta M_i^{local}-2h\sigma_i m \le -2h/N$; and this is a stronger
condition for a solution than $\Delta M_i^{local}-2h\sigma_i m \le 0$ when $h>0$. Indeed, there is a finite probability to miss a solution by adding the small
term $2h/N$ even in the large $N$ limit. Secondly, the size of population that we use to represent the statistics of BP messages is finite. It means that, even
if in the thermodynamic limit we do not have any solution, we may still observe a positive entropy
due to the finite size of the population. What we can do is just to work with
the largest population allowed in our numerical simulations.

The local problem can also be used to converge the system to a problem solution by applying reinforcement. Here are the reinforced BP equations

\begin{eqnarray}
\pi_{i \rightarrow j}(\sigma_i;\sigma_j)\propto [\pi_{i}(\sigma_i)]^r
 \sum_{\{\sigma_k|k\in \partial i \setminus j\}} I_i(\sigma_i|\sigma_{\partial i},m) \prod_{k \in \partial i \setminus j}\pi_{k \rightarrow i}(\sigma_k;\sigma_i),\\ \nonumber
\pi_{i}(\sigma_i)\propto [\pi_{i}(\sigma_i)]^r
\sum_{\{\sigma_k|k\in \partial i\}} I_i(\sigma_i|\sigma_{\partial i},m) \prod_{k \in \partial i}\pi_{k \rightarrow i}(\sigma_k;\sigma_i).
\end{eqnarray}

\begin{figure}
\includegraphics[width=10cm]{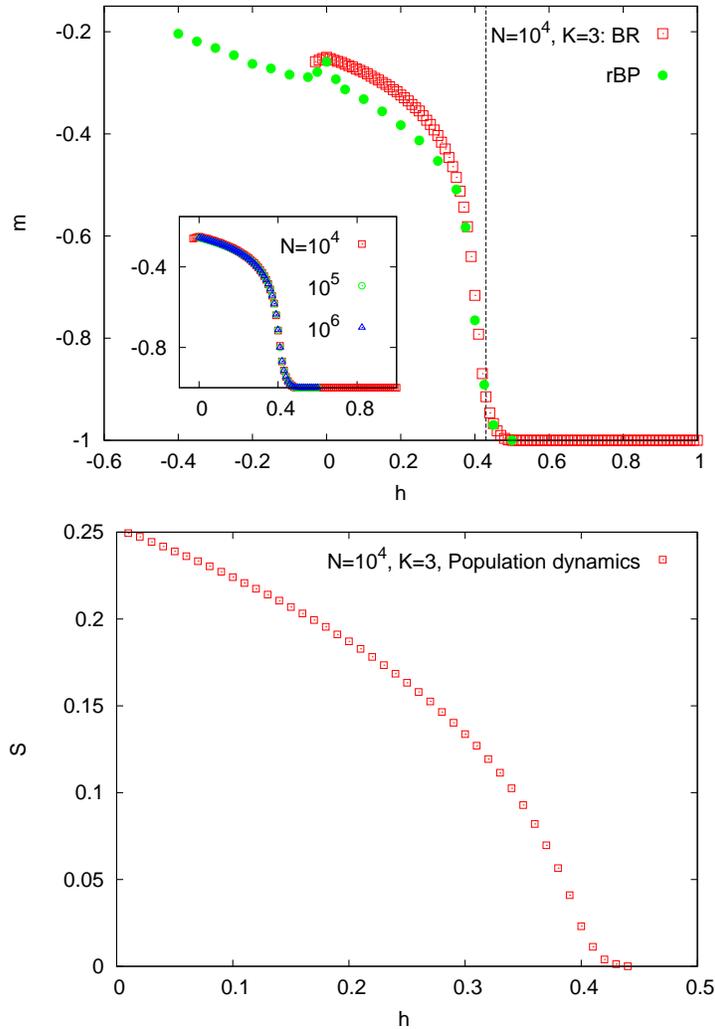}
\caption{ Upper plot: average magnetization of solutions found by the rBP and BR algorithms in $100$ instances of the global problem with local mIS constraints.
The vertical line shows the point where globally dominated solutions appear. The inset displays the behavior for different sizes. Lower plot: typical entropy computed with population dynamics.}\label{f11}
\end{figure}

The algorithm works by computing $m=\frac{1}{N}\sum_i [\pi_i(+1)-\pi_i(-1)]$ from the BP marginals at each iteration and using that as an estimation of total magnetization in the
Nash conditions. Remember that $m$ can be computed locally by passing appropriate messages $g_{i\to j}$ along a spanning tree.
Figure \ref{f11} compares the magnetization of solutions found in this way with the Best Response algorithm in a problem with local mIS constraints. We see that even for a large graph of $N=10^4$ players, there are still some Nash equilibria for $0<2hm<1$, where the solution set is asymptotically empty. 
The Best Response algorithm finds easily a solution when the global field is positive but it does not converge for negative $h$. 
Using the above rBP algorithm we could obtain different kinds of solutions depending on the reinforcement parameter. 
The solutions lay, of course, in the region that BP equations converge. In figure \ref{f11} we also show the typical entropy for positive values of the global field.  
It seems that both the entropy and magnetization approach continuously to their limiting global values.

\section{Conclusion}\label{S5}
We used the cavity method in conjunction with rigorous bounds to study random graphical games with local and global interactions. 
We analyzed  the phase diagram of the problem and presented some  local message passing algorithms  which allow to find efficiently a Nash equilibrium.

More specifically, we studied graphical games with local payoffs coming from the maximal independent set problem and global payoffs which depend on the average strategy  over the whole graph (the so called total magnetization in the physics jargon). Introducing the global interaction resulted to a new set of equilibria which are a mixture of locally happy and unhappy players. In summary: 
(i) Using rigorous arguments and the annealed entropy, we showed that these equilibria cannot be present in some regions of the phase space in the thermodynamic limit.
(ii) We observed an exponentially large number of these equilibria for positive global fields and negative magnetizations which we conjecture survive in the thermodynamic limit.
Indeed, a simple heuristic algorithm like the Best Response is able to find such a typical solution in very large problem instances. The entropy and the total magnetization of the typical equilibria decrease continuously as we increase the global field starting from zero. 
(iii) Numerical simulations and annealed approximation results show the existence of a critical value of $h$ above which a cluster of solutions dominated by the global interaction appears.     
Close to this value of the global field, the typical entropy and magnetization decrease very rapidly, separating the regions governed by the local and global interactions.

\acknowledgments
We would like to thank A. Braunstein, L. Dall'Asta  and M. Marsili for useful discussions.

\appendix

\section{More details on Remarks $(1)$ to $(4)$}\label{remarks}

\textbf{On Remark (1)}:
Consider a solution of the problem with magnetization $m$ and global field $h$.
In a local solution players are either in set $H_+$ or $H_-$.  For a plus player the Nash condition is $\Delta M^{local}-2h(m-1/N)<0$ which for happy players means
$-1-2h(m-1/N)<0$. Notice that for integer payoffs in $\{0,1\}$ a happy player has $\Delta M^{local}=-1$ whereas for an unhappy player $\Delta M^{local}=+1$.
Therefore, a happy plus player is plying the best response as long as

\begin{equation}
\left\{ \begin{array}{ll}
    m-\frac{1}{N} < \frac{1}{2|h|}, & \hbox{$h<0$;} \\
    m-\frac{1}{N} > -\frac{1}{2|h|}, & \hbox{$h>0$;} \\
  \end{array}
\right.
\end{equation}

For a happy minus player the Nash condition is $-1+2h(m+1/N)<0$  which is satisfied when

\begin{equation}
\left\{ \begin{array}{ll}
    m+\frac{1}{N} > -\frac{1}{2|h|}, & \hbox{$h<0$;} \\
    m+\frac{1}{N} < \frac{1}{2|h|}, & \hbox{$h>0$;} \\
  \end{array}
\right.
\end{equation}

To have a local solution we just need $-m^*(h)<m<m^*(h)$ where $m^*(h)=\min(1,\frac{1}{2|h|}-\frac{\mathrm{sgn}(h)}{N})$. If $|h|<\frac{1}{2(1+\mathrm{sgn}(h)/N)}$ we get $m^*=1$, that is all local solutions survive
after adding the global incentives.

\textbf{On Remark (2)}:
According to the above arguments, to have both happy plus and minus players when $h>N/2$ we need at the same time $m>0$ and $m<0$, which is impossible. But
we can have all plus or all minus solutions, that is the two global solutions for $h>0$.  Indeed these two solutions appear at $h=\frac{1}{2(1 - 1/N)}$ where unhappy players are allowed.
To see this we note that a player in $U_+$ is satisfied if

\begin{equation}
\left\{ \begin{array}{ll}
    m-\frac{1}{N} < -\frac{1}{2|h|}, & \hbox{$h<0$;} \\
    m-\frac{1}{N} > \frac{1}{2|h|}, & \hbox{$h>0$;} \\
  \end{array}
\right.
\end{equation}

The all plus solution is possible if $1>\frac{1}{2h}+\frac{1}{N}$, that is $h>\frac{1}{2}+\frac{h}{N}$.
For players in $U_-$ we have

\begin{equation}
\left\{ \begin{array}{ll}
    m+\frac{1}{N} > \frac{1}{2|h|}, & \hbox{$h<0$;} \\
    m+\frac{1}{N} < -\frac{1}{2|h|}, & \hbox{$h>0$;} \\
  \end{array}
\right.
\end{equation}

The all minus solution is possible if $-1<-\frac{1}{2h}-\frac{1}{N}$, that is $h>\frac{1}{2(1-1/N)}$.

On the other hand, to have the $m=0$ global solutions for $h<0$, we need $0>\frac{1}{2|h|}-\frac{1}{N}$ which implies $|h|>N/2$.

\textbf{On Remark (3)}:
From the above equations we see that to have players in both $U_+$ and $U_-$ when $h>0$, we need $m > \frac{1}{2|h|}+\frac{1}{N}$ and $m < -\frac{1}{2|h|}-\frac{1}{N}$ which is not possible. Indeed one could have only unhappy plus players if $m>0$ or unhappy minus players if $m<0$. But a mixed solution should contain both plus and minus players, otherwise it would be completely polarized global solution.
Considering the case $m>0$, the conditions for unhappy plus players and happy minus players give contradictory inequalities $m>\frac{1}{2|h|}+\frac{1}{N}$ and $m<\frac{1}{2|h|}-\frac{1}{N}$.
In the case $m<0$, the conditions for unhappy minus players and happy plus players give contradictory inequalities $m>-\frac{1}{2|h|}+\frac{1}{N}$ and $m<-\frac{1}{2|h|}-\frac{1}{N}$. Therefore, we can not have a mixed solution for positive global fields.

When $h<0$, to have non-empty sets $U_+$ and $U_-$ we need $m < -\frac{1}{2|h|}+\frac{1}{N}$ and
$m > \frac{1}{2|h|}-\frac{1}{N}$. Again we can only unhappy plus players if $m<0$ or unhappy minus players if $m>0$. In other words, we can only have solutions of type $(H_+,H_-,U_+)$ if $hm>0$ and
$(H_+,H_-,U_-)$ if $hm<0$. Considering the two cases $m>0$ and $m<0$ we find $\frac{1}{2|h|}-\frac{1}{N}<m<\frac{1}{2|h|}+\frac{1}{N}$ and $-\frac{1}{2|h|}-\frac{1}{N}<m<-\frac{1}{2|h|}+\frac{1}{N}$, respectively. That is,
if there is a mixed solution it should have magnetization $\pm \frac{1}{2|h|}$ and so $2|h|>1$.

Note also that for both $h>0$ and $h<0$ we can not have an $m=0$ solution of type $(H_+,H_-,U_+,U_-)$ as long as $|h|<N/2$.

\textbf{On Remark (4)}:
Consider the maximal independent set problem on graph $\mathcal{G}$ and let $\rho_{max}$ be the size of
a maximum independent set. Thus, all local solutions have a magnetization less than $2\rho_{max}-1$.
Consider the global problem when $h>0$. According to the previous remarks, if there is a solution of magnetization $2\rho_{max}-1<m<0$ to the problem, it should be of type $(H_+,H_-,U_-)$. But this is not possible because we can always change the state of a player in $U_-$ to $+1$, increasing the magnetization and still respecting the mIS constraints.
In other words, such a configuration would lead to local solution of magnetization $m>2\rho_{max}-1$, which is of course a contradiction.

The other part of remark (4) can be proved by
bounding the probability of having a solution, as we did in section \ref{S33} for random payoffs.
Consider a region $\Omega$ which is a chain of three neighboring players $(j,i,k)$. Here we show that if $0<2h(m \pm 1/N)<1$, then for any
boundary configuration $\sigma_{\partial \Omega}$ there is a nonzero probability of having no solution in the region.

Let us start by eliminating the two
configurations $(+1,+1,-1),(-1,+1,+1)$ in which player $i$ is always locally unhappy, independent of the boundary configuration. To do this we need
$|\Delta M_i^{local}(\sigma_i,\sigma_{\partial i})|> 2h(m-1/N)$ that could happen with a nonzero probability as long as $2h(m-1/N)<1$. Now consider configuration $(+1,-1,+1)$ where
player $i$ is always locally happy. Again we can avoid this solution by choosing $|\Delta M_i^{local}(\sigma_i,\sigma_{\partial i})|< 2h(m+1/N)$, which is possible if $2h(m+1/N)>0$.

Then consider three configurations $(-1,+1,-1),(-1,-1,+1),(-1,-1,-1)$ where player $j$ could be locally happy or unhappy.
In each case we can eliminate the solution by choosing $|\Delta M_j^{local}(\sigma_j,\sigma_{\partial j})|< 2h(m+1/N)$ or $|\Delta M_j^{local}(\sigma_j,\sigma_{\partial j})|> -2h(m+1/N)$. This can be done with a nonzero probability if $2h(m+1/N)>0$.

There remain two configurations $(+1,-1,-1)$ and $(+1,+1,+1)$ where player $k$ could be locally happy or unhappy. Again we can
choose $|\Delta M_k^{local}(\sigma_k,\sigma_{\partial k})|< 2h(m+1/N)$ or $|\Delta M_k^{local}(\sigma_k,\sigma_{\partial k})|> -2h(m+1/N)$ to eliminate $(+1,-1,-1)$ if $2h(m+1/N)>0$. And we choose $|\Delta M_k^{local}(\sigma_k,\sigma_{\partial k})|> 2h(m-1/N)$  to eliminate $(+1,+1,+1)$ if $2h(m+1/N)<1$. Therefore, we have a nonzero probability $P_{no solution}(\Omega)$ which in the thermodynamics limit gives an exponentially small probability of having solution  $P_{solution}< (1-P_{no solution}(\Omega))^{N_{\Omega}}$.

\end{document}